\documentclass[conference]{IEEEtran}

\usepackage{color}

\usepackage{cite}

\ifCLASSINFOpdf
 \usepackage[pdftex]{graphicx}
\else
   \usepackage[dvips]{graphicx}
\fi

\usepackage[cmex10]{amsmath}
\usepackage{amsfonts,amssymb,bbm,array,amsthm}

\usepackage{array}

\usepackage{multirow}

\newcommand{\superscript}[1]{\ensuremath{^{\textrm{#1}}}} % e.g. Welcome to the 21\superscript{st} century.

\begin{document}
\title{Improved Channel Estimation Methods based on PN sequence for TDS-OFDM}

\author{
\IEEEauthorblockN{Ming Liu, Matthieu Crussi\`ere, Jean-Fran\c{c}ois H\'elard}
\IEEEauthorblockA{Universit\'e Europ\'eenne de Bretagne (UEB)\\
INSA, IETR, UMR 6164, F-35708, Rennes, France \\
Email: \{ming.liu; matthieu.crussiere; jean-francois.helard\}@insa-rennes.fr}
}

\maketitle

\begin{abstract}
An accurate channel estimation is crucial for the novel time domain synchronous orthogonal frequency-division multiplexing (TDS-OFDM) scheme in which pseudo noise (PN) sequences serve as both guard intervals (GI) for OFDM data symbols and training sequences for synchronization/channel estimation. This paper studies the channel estimation method based on the cross-correlation of  PN sequences. A theoretical analysis of this estimator is conducted and several improved estimators are then proposed to reduce the estimation error floor encountered by the PN-correlation-based estimator. It is shown through mathematical derivations and simulations that the new estimators approach or even achieve the Cram\'er-Rao bound.
\end{abstract}

\section{Introduction}
\IEEEPARstart{I}{n} the recently proposed time domain synchronous orthogonal frequency-division multiplexing (TDS-OFDM) scheme~\cite{wang2005iterative}, the classical cyclic prefix (CP) conventionally used in OFDM is replaced by a known pseudo noise sequence (simply termed as PN hereafter) which is reused as training sequence for channel estimation and synchronization.
Consequently, TDS-OFDM combines the guard interval (GI) and the training symbols and does not need any additional pilots in the frequency domain, thereby achieving a higher spectral efficiency than CP-OFDM.
TDS-OFDM has been adopted by the novel Chinese digital television broadcasting standard--DTMB~\cite{DTMB_Standard}.

In TDS-OFDM, a channel estimate is needed to separate the PN from the OFDM data part at the receiver.
Its accuracy is crucial for the demodulation process to avoid any residual PN components in the received signal.
Hence, channel estimation plays a prominent part in TDS-OFDM performance and needs to be carefully studied.
\cite{ma2006channel} investigates several channel estimation techniques based on known sequences but did not exploit the property of the PN.
\cite{wang2005iterative} uses the cross-correlation results of the transmitted and received PNs as the channel estimates ignoring the fact that the correlation of the PN is not a perfect impulse function, which introduces mutual interference between different channel paths in the estimation result.
To overcome this problem, \cite{song2005channel} suggests to remove interference components of several strongest paths and \cite{Liu2007ITD} proposes to iteratively detect the significant paths in the correlation results.
Both of them did not thoroughly eliminate the interference lying in the PN correlation.
Finally, \cite{Xi2005Fast} proposes a least square (LS) channel estimator based on the property of maximum length sequence (m-sequence). This solution however gives a suboptimal performance as explained in the sequel.

In this paper, we investigate the performance of the PN-correlation-based channel estimator, and, more importantly, propose several improved estimators that reduce or even eliminate the estimation error floor resulting from the interference term issued from the PN correlation function.
In the following, section II introduces the PN based channel estimation and gives a theoretical analysis of its performance. In section III, three different improved estimators are proposed and their respective mean square errors (MSE) are derived. Finally, their performance is compared in section IV through simulations over different channel conditions.

\section{PN-Correlation-based Channel Estimator}
\begin{figure}[!t]
\centering
\includegraphics[width=3in]{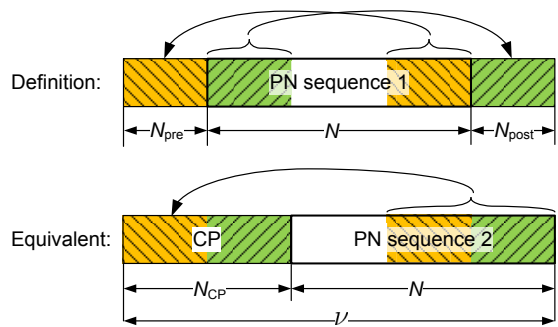}
\caption{Structure of the GI specified in the DTMB standard.}
\label{fig_PN}
\end{figure}

\subsection{System Model}
In the DTMB system, the $\nu$-length GI consists of an $N$-length m-sequence as well as its pre- and post- circular extensions~\cite{DTMB_Standard}.
Since any circular shift of an m-sequence is itself an m-sequence~\cite{Sarwate1980Crosscorrelation}, the GI can also be treated as another $N$-length m-sequence, denoted by $\textbf{p}=[p_1,p_2,\ldots, p_N]^T$ with $(\cdot)^T$ standing for the matrix transpose, and an ${N_{_{\textrm{CP}}}}$-length CP.
That is, if this CP is longer than the channel delay spread, the CP absorbs the channel time dispersions and the $N$-length m-sequence is ISI-free.
The received PN can be written as:
\begin{equation}
\label{eqn_pn_receive}
\textbf{d} = \textbf{H} \textbf{p} + \textbf{w} = \textbf{P} \textbf{h} + \textbf{w},
\end{equation}
where $\textbf{d}=[d_1,d_2,\ldots, d_N]^T$, $\textbf{w}=[w_1,w_2,\ldots,w_N]^T$ and  $\textbf{h}=[h_1,h_2,\ldots, h_L,0,\ldots, 0]^T$
are the received PN, additive white Gaussian noise (AWGN), and channel impulse response (CIR), respectively.
The channel is modeled as an $L$\superscript{th}-order finite impulse response (FIR) filter with
$\alpha_{l}^2=\mathbb{E}\big[ |h_l|^2 \big]$ the average power of the $l$\superscript{th} channel tap.
The power of the channel paths is normalized such that $\sum_{l=1}^{L}\alpha_{l}^2=1$.
$\textbf{H}$ and $\textbf{P}$ are the $N\times N$ circulant matrices with first rows $[h_1,0,\ldots, 0,h_{L},\ldots, h_2]$ and $[p_1,p_N,p_{N-1},\ldots, p_2]$, respectively. The second equality in (\ref{eqn_pn_receive}) uses the commutativity of the convolution.

\subsection{PN Correlation-based Estimator}
Channel estimation can be simply obtained by performing time domain correlation of transmitted and received PN sequences.
Recall the circular autocorrelation property of the m-sequence~\cite{Sarwate1980Crosscorrelation}:
\begin{eqnarray}
  \label{eqn_property_PN}
  R(n)=\frac{1}{N}\sum_{m=1}^{N}p_m^{}p_{[m+n]_N}^{\ast}= \left\{ {\begin{array}{*{20}c}{ 1 } & {n = 0} \\
   { - \frac{1}{N}} & {0 < n < N}  \\ \end{array}} \right. ,
\end{eqnarray}
where $(\cdot)^{\ast}$ is the conjugate of complex number and $[\cdot]_N$ denotes modulo-$N$ operation.
When the PN is sufficiently long, i.e., $- \frac{1}{N}\approx0$, the autocorrelation function approaches the \textit{Kronecker} delta function.
The CIR can be extracted from the received PN using this delta function.
The estimator is:
\begin{eqnarray}
\label{eqn_bar_h_mat}
  &&\bar{\textbf{h}}=\frac{1}{N}\textbf{C}\textbf{d} =  \frac{1}{N}\textbf{C}\textbf{P}\textbf{h} + \frac{1}{N}\textbf{C}\textbf{w},
\end{eqnarray}
where $\textbf{C}=\textbf{P}^{\mathcal{H}}$ is an $N\times N$ circulant matrix with first row $\big [p_1,p_2,\ldots,p_N\big]$ and represents the circular correlation of the PN.
$(\cdot)^{\mathcal{H}}$ is the Hermitian transpose.
Using the correlation property (\ref{eqn_property_PN}), it yields: %$\textbf{Q}\triangleq{ \frac{1}{N}\textbf{C}\textbf{P}}$.

\begin{eqnarray}
 &&\textbf{Q}\triangleq{ \frac{1}{N}\textbf{C}\textbf{P}} = \left[ \begin{array}{*{20}c}
   1 & -\frac{1}{N}&   \cdots  & -\frac{1}{N} \\
    -\frac{1}{N} &  1  &  \cdots  &  -\frac{1}{N}  \\
    \vdots  &  \vdots  &    \ddots  &  \vdots   \\
    -\frac{1}{N}&  -\frac{1}{N}&   \cdots  & 1  \\
\end{array} \right].
\end{eqnarray}
The estimation error is $\boldsymbol \xi_{\bar{h}} = \bar{\textbf{h}} - \textbf{h}=(\textbf{Q}-\textbf{I}_{N})\textbf{h}+ \frac{1}{N}\textbf{C}\textbf{w}$, where $\textbf{I}_{N}$ is an $N\times N$ identity matrix.
Assuming that the channel taps are uncorrelated, the MSE of the estimate is:
\begin{eqnarray}
&&\varepsilon_{\bar{h}}=\frac{1}{N}\mathrm{Tr}\left(\mathbb{E}\big[\boldsymbol \xi_{\bar{h}} \boldsymbol \xi_{\bar{h}}^{\mathcal{H}}\big] \right)\nonumber\\
&=& \!\!\!\!\!\!\frac{1}{N}\mathrm{Tr}\!\left(\mathbb{E}\big[(\!\textbf{Q}\!-\!\textbf{I}_{N}\!)\textbf{h} \textbf{h}^{\mathcal{H}}\!(\!\textbf{Q}\!-\!\textbf{I}_{N}\!)^{\!\mathcal{H}} \big] \right)\!\!+\!\!  \frac{1}{N^3}\mathrm{Tr}\!\left(\mathbb{E}\big[ \textbf{C}\textbf{w}\textbf{w}^{\mathcal{H}} \textbf{C}^{\mathcal{H}} \big] \right)\nonumber\\
&=&\!\!\!\!\!\! \underbrace{\frac{1}{N}\mathrm{Tr}\left((\textbf{Q}-\textbf{I}_{N})\boldsymbol \Lambda  (\textbf{Q}-\textbf{I}_{N})^{\mathcal{H}} \right)}_{{\rm{interference}}} +  \underbrace{\frac{\sigma^2_w}{N^3}\mathrm{Tr}\left( \textbf{C} \textbf{C}^{\mathcal{H}} \right)}_{\rm noise},
\end{eqnarray}
where $\mathrm{Tr}(\cdot)$ is the trace of a matrix, $\boldsymbol \Lambda$ is a diagonal matrix of size $N$ whose first $L$ elements of the main diagonal is $\big[ \alpha_1^2,\alpha_2^2,\ldots,\alpha_{L}^2 \big]$, and the rest elements are all $0$'s.
The estimation error is composed of two parts: the interference resulting from the correlation of PN sequences and the noise.
It can be found that the interference comes from the contributions of the off-diagonal elements in $\textbf{Q}$ and
 vanishes if $\textbf{Q}=\textbf{I}_{N}$.
The estimator is asymptotically unbiased as $N \to \infty $.
More concretely, the interference is computed as:
\begin{eqnarray}
&&(\textbf{Q}-\textbf{I}_{N})\boldsymbol \Lambda  (\textbf{Q}-\textbf{I}_{N})^{\mathcal{H}}=\nonumber\\
  &&
    \left[ \begin{smallmatrix}
   {{\textstyle{(1 - \alpha_1^2 ) \over {N^2 }}}} & {{\textstyle{(1 - \alpha _2^2  - \alpha_1^2 ) \over {N^2 }}}} &  \cdots  & {{\textstyle{(1 - \alpha_{\scriptscriptstyle L}^2  - \alpha _1^2 ) \over {N^2 }}}} &  &  \cdots  &   \\
   {{\textstyle{(1 - \alpha_1^2  - \alpha _2^2 ) \over {N^2 }}}} & {{\textstyle{(1 - \alpha_2^2 ) \over {N^2 }}}} &  \cdots  & {{\textstyle{(1 - \alpha _{\scriptscriptstyle L}^2  - \alpha _2^2 ) \over {N^2 }}}} &  &  &   \\
    \vdots  &  \vdots  &  \ddots  &  \vdots  &    & {} &     \\
   {{\textstyle{(1 - \alpha _1^2  - \alpha _{\scriptscriptstyle L}^2 ) \over {N^2 }}}} & {{\textstyle{(1 - \alpha _2^2  - \alpha _{\scriptscriptstyle L}^2 ) \over {N^2 }}}} &  \cdots  & {{\textstyle{(1 - \alpha _{\scriptscriptstyle L}^2 ) \over {N^2 }}}} &  & \ldots &   \\
    &  & {} &  & {\textstyle{1 \over {N^2 }}} & {} &   \\
    \vdots  &  \vdots  & {} &  \vdots  & {} &  \ddots  &     \\
    &  &    &  &  &    & {\textstyle{1 \over {N^2 }}}  \\
 \end{smallmatrix} \right].
\end{eqnarray}
Therefore, the MSE of the estimator is finally:
\begin{equation}
\label{eqn_mse_bar_h_time}
  \varepsilon_{\bar{h}} = \frac{\sigma^2_w}{N} + \frac{N-1}{N^3}.
\end{equation}
Eventually, the first term of the MSE expression is proportional to the noise variance, while the second term is only determined by the length of the PN.
In other words, it produces an estimation error floor with an MSE level of ${(N-1)}/N^3$.
This error floor appears when ${(N-1)}/{N^3}>{\sigma^2_w}/{N}$, i.e., $\mathrm{SNR} (\mathrm{dB})>10\log_{10}{({N^2 }/(N-1) )}$.
For example, the error floor appears when the SNR is greater than $24.1$~dB and $27.1$~dB, given the $255$-length and $511$-length PNs specified in~\cite{DTMB_Standard}, respectively.

The Cram\'er-Rao bound of training sequence based channel estimation with a length equal to the training sequence is~\cite{de1997cramer}:
\begin{equation}
\label{eqn_cramer_rao}
  MSE\geq\frac{\sigma_w^2}{N}\mathrm{Tr}\left((\textbf{P}^{\mathcal{H}}\textbf{P})^{-1}\right)=\frac{\sigma^2_w}{N+1}.
\end{equation}

Comparing (\ref{eqn_mse_bar_h_time}) and (\ref{eqn_cramer_rao}), it can be found that the correlation based estimator approaches the Cram\'er-Rao bound at low SNR (i.e. $\sigma^2_w$ is large), but suffers an estimation error floor at high SNR. Therefore, we go in for some improved estimators aiming at reducing this error floor.

\section{Improved Estimators with Reduced Error Floor}
\label{section:improved_time_est}

\subsection{Method 1: Multiplying by Inverse of Matrix $\textbf{Q}$}
From the analysis in the last section, the estimation error floor comes from the fact that $\textbf{Q}$ is not a perfect identity matrix.
Therefore, a straightforward solution is to perform a linear transformation $\boldsymbol \Omega$ such that $\boldsymbol \Omega \textbf{Q}=\textbf{I}_{N}$.
Since $\textbf{Q}$ is known and always full rank for a given m-sequence, $\boldsymbol \Omega = \textbf{Q}^{-1}$.
A new estimator is obtained by left multiplying the correlation-based estimator (\ref{eqn_bar_h_mat}) by $\textbf{Q}^{-1}$:
\begin{equation}
\label{eqn_est_mult_q_inv}
  \hat{\textbf{h}}_1=\textbf{Q}^{-1}\bar{\textbf{h}}=\textbf{h}+\frac{1}{{N}}\textbf{Q}^{-1}\textbf{C}\textbf{w},
\end{equation}
which leads to an LS estimator that is in some extend similar to that proposed in \cite{Xi2005Fast}.
The estimation error is $\boldsymbol \xi_{\hat{h}_1}=\frac{1}{{N}}\textbf{Q}^{-1}\textbf{C}\textbf{w}$.
The MSE of the estimator is:
\begin{eqnarray}
\label{eqn_mse_hat_h_time_1}
  &&\varepsilon_{\hat{h}_1}=\frac{1}{N}\mathrm{Tr}\left(\mathbb{E}\big[\boldsymbol \xi_{\hat{h}_1} \boldsymbol \xi_{\hat{h}_1}^{\mathcal{H}}\big] \right)\nonumber\\
&=&\frac{\sigma_w^2}{N^3}\mathrm{Tr}\left( \textbf{Q}^{-1}\textbf{C}\textbf{C}^{\mathcal{H}} (\textbf{Q}^{-1})^{\mathcal{H}} \right)%\nonumber\\
=\frac{\sigma_w^2}{N^2}\mathrm{Tr}\left(\textbf{Q}^{-1}\right).
\end{eqnarray}
As $\textbf{Q}$ is a circulant matrix and all its elements are known for a given m-sequence, its inverse is easily obtained and does not need complex computations~\cite{Gray2006Toeplitz}:
\begin{equation}
\label{eqn_cor_inv}
  \textbf{Q}^{-1}=\left[ {\begin{array}{*{20}c}
   a & b &   \cdots  & b \\
   b & a &   \cdots  & b \\
    \vdots  &  \vdots  &   \ddots  &  \vdots   \\
   b & b &\cdots  & a \\
\end{array}} \right]
\end{equation}
where
\begin{equation}
  a = \frac{2N}{N+1},\ \mathrm{and}\ b=\frac{N}{N+1}. \nonumber
\end{equation}
Replacing (\ref{eqn_cor_inv}) into (\ref{eqn_mse_hat_h_time_1}), the MSE becomes:
\begin{equation}
\label{eqn_mse_hat_h_time_1_final}
  \varepsilon_{\hat{h}_1} = \frac{2\sigma_w^2}{N+1}.
\end{equation}
Comparing (\ref{eqn_mse_hat_h_time_1_final}) and (\ref{eqn_mse_bar_h_time}), it can be found that the estimation error floor is removed by left multiplying by $\textbf{Q}^{-1}$. In the meantime, the MSE is however approximately twice as much as the Cram\'er-Rao bound due to a noise power increase.

\subsection{Method 2: Multiplying by Inverse of Truncated Matrix $\bar{\textbf{Q}}$}
Suppose the length of the CIR $L$ is perfectly known, the estimate (\ref{eqn_bar_h_mat}) is truncated to $L$-length. The estimator becomes:
\begin{equation}
\label{eqn_h_est_trunc_mat}
  \tilde{\textbf{h}} = \textbf{T}\bar{\textbf{h}} =\bar{\textbf{Q}}\textbf{h} + \frac{1}{N}\textbf{T}\textbf{C}\textbf{w},
\end{equation}
where $\textbf{h}$ is the $L$-length vector of the real CIR. $\textbf{T}$ is an $L\times N$ matrix whose left $L\times L$ submatrix is an identity matrix  and rest parts are all $0$'s, which represents deleting the last $N-L$ elements of an $N$-length vector. $\bar{\textbf{Q}}\triangleq\textbf{T}\textbf{Q}\textbf{T}^{T}$ is an $L\times L$ circulant matrix that contains the first $L$ rows and $L$ columns of matrix $\textbf{Q}$ and is still full rank.
Similarly to (\ref{eqn_est_mult_q_inv}), we can left multiply estimator (\ref{eqn_h_est_trunc_mat}) by the inverse matrix $\bar{\textbf{Q}}^{-1}$ to obtain:
\begin{equation}
\label{eqn_est_q_inv_trunc}
  \hat{\textbf{h}}_2=\bar{\textbf{Q}}^{-1}\tilde{\textbf{h}}=\textbf{h}+\frac{1}{{N}}\bar{\textbf{Q}}^{-1}\textbf{T}\textbf{C}\textbf{w}.
\end{equation}
This new estimator is still unbiased. The related estimation error is $\boldsymbol \xi_{\hat{h}_{2}}=\frac{1}{{N}}\bar{\textbf{Q}}^{-1}\textbf{T}\textbf{C}\textbf{w}$ which leads to an MSE given by:
\begin{eqnarray}
\label{eqn_mse_hat_h_time_1_trunc}
  &&\varepsilon_{\hat{h}_2}=\frac{1}{L}\mathrm{Tr}\left(\mathbb{E}\big[\boldsymbol \xi_{\hat{h}_2} \boldsymbol \xi_{\hat{h}_2}^{\mathcal{H}}\big] \right)\nonumber\\
&=&\frac{\sigma_w^2}{LN}\mathrm{Tr}\left(\bar{\textbf{Q}}^{-1}\bar{\textbf{Q}} (\bar{\textbf{Q}}^{-1})^{\mathcal{H}}\right)%\nonumber\\
=\frac{\sigma_w^2}{LN}\mathrm{Tr}\left(\bar{\textbf{Q}}^{-1}\right).
\end{eqnarray}
The inverse matrix $\bar{\textbf{Q}}^{-1}$ is the $L\times L$ circulant matrix with a similar form as (\ref{eqn_cor_inv}). The elements $a$ and $b$ are replaced by
\begin{eqnarray}
  \bar{a} &=& 1+\frac{L-1}{N^2+2N-NL-L+1},\ \mathrm{and} \nonumber\\
  \bar{b} &=& \frac{N}{N^2+2N-NL-L+1}. \nonumber
\end{eqnarray}

The MSE of the estimation is finally:
\begin{equation}
\label{eqn_mse_hat_h_time_1_trunc_final}
  \varepsilon_{\hat{h}_2}=\frac{N-L+2}{N^2+2N-NL-L+1}\sigma_w^2.
\end{equation}
Comparing (\ref{eqn_mse_hat_h_time_1_final}) and (\ref{eqn_mse_hat_h_time_1_trunc_final}), it can be found that the estimation MSE is reduced thanks to the truncation process. In the extreme case when $N=L$, the estimator (\ref{eqn_est_q_inv_trunc}) turns to (\ref{eqn_est_mult_q_inv}).

When the channel length $L$ is known, the Cram\'er-Rao bound of a training sequence based channel estimation is computed by replacing matrix $\textbf{P}$ by $\bar{\textbf{P}}=\textbf{P}\textbf{T}^{T}$ in (\ref{eqn_cramer_rao}):
\begin{equation}
  MSE\!\geq\!\!\frac{\sigma_w^2}{L}\mathrm{Tr}\!\left(\!(\bar{\textbf{P}}^{\mathcal{H}}\bar{\textbf{P}})^{-1}\!\right)\!=\!\frac{(N-L+2)\sigma^2_w}{N^2+2N-NL-L+1}.
\end{equation}
This demonstrates that the proposed estimator (\ref{eqn_mse_hat_h_time_1_trunc}) achieves the Cram\'er-Rao bound.

\subsection{Method 3: Subtracting Interference}

Another improved estimator can be obtained by \textit{subtracting} the contribution of the interference from the correlation-based estimator.

More precisely, suppose the CIR length $L$ is known, we can rewrite the truncated CIR estimate (\ref{eqn_h_est_trunc_mat}) as:
\begin{equation}
  \tilde{\textbf{h}} = \textbf{h} + \bar{\boldsymbol \Delta} \textbf{h} +\frac{1}{N}\textbf{T}\textbf{C}\textbf{w},
\end{equation}
where $\bar{\boldsymbol \Delta}$ is the $L\times L$ matrix that contains all the off-diagonal elements of $\bar{\textbf{Q}}$  and the main diagonal elements of $\bar{\boldsymbol \Delta}$ are $0$'s, $\textbf{h}$ is the $L$-length vector of the real CIR.
Considering that $\bar{\boldsymbol \Delta}$ is known for a given m-sequence, we propose to use the estimated $\tilde{\textbf{h}}$ to reduce the interference. An estimator is thus:
\begin{equation}
\label{eqn_h_est_subract}
  \hat{\textbf{h}}_3=\tilde{\textbf{h}}-\bar{\boldsymbol \Delta}\tilde{\textbf{h}}
   =\textbf{h} - \bar{\bar{\boldsymbol \Delta}}\textbf{h}  + \frac{1}{N}\textbf{T}\textbf{C}\textbf{w} -\frac{1}{N}\bar{\boldsymbol \Delta}\textbf{T}\textbf{C}\textbf{w},
\end{equation}
where
\begin{equation}
\bar{\bar{\boldsymbol \Delta}} \triangleq  \bar{\boldsymbol \Delta}\bar{\boldsymbol \Delta}= \left[ \begin{array}{*{20}c}
    \frac{L-1}{N^2} & \frac{L-2}{N^2}&  \cdots  &\frac{L-2}{N^2} \\
    \frac{L-2}{N^2} &   \frac{L-1}{N^2} &  \cdots  & \frac{L-2}{N^2} \\
    \vdots  &  \vdots   &  \ddots  &  \vdots   \\
    \frac{L-2}{N^2}&  \frac{L-2}{N^2}&  \cdots  &  \frac{L-1}{N^2}   \\
\end{array} \right].
\end{equation}
The estimation error is: $\boldsymbol \xi_{\hat{h}_3}= - \bar{\bar{\boldsymbol \Delta}}\textbf{h} + \frac{1}{N}\textbf{T}\textbf{C}\textbf{w}-\frac{1}{N}\bar{\boldsymbol \Delta}\textbf{T}\textbf{C}\textbf{w}.$
Recalling that $\bar{\textbf{Q}}=\frac{1}{N}\textbf{T}^{}\textbf{C}^{}\textbf{C}^{\mathcal{H}}\textbf{T}^{\mathcal{H}}$, the MSE of the estimate is:
\begin{eqnarray}
\label{eqn_mse_h2_general}
  &&\varepsilon_{\hat{h}_3}=\frac{1}{L}\mathrm{Tr}\left(\mathbb{E}\big[\boldsymbol \xi_{\hat{h}_3} \boldsymbol \xi_{\hat{h}_3}^{\mathcal{H}}\big] \right)\nonumber\\
&=&\!\!\!\!\!  \frac{1}{L}\mathrm{Tr}\!\left(\!\! \bar{\bar{\boldsymbol \Delta}} \bar{\boldsymbol \Lambda} (\bar{\bar{\boldsymbol \Delta}})^{\mathcal{H}}\!
\!+\! \frac{\sigma_w^2}{N}\!\Big(\!\bar{\textbf{Q}}
\!+\!\bar{\boldsymbol \Delta}\bar{\textbf{Q}}\bar{\boldsymbol \Delta}^{\!\mathcal{H}}\!\!
\!-\!2\Re\big\{\bar{\textbf{Q}}\bar{\boldsymbol \Delta}^{\!\mathcal{H}}\!\big\}\!\Big)\!\!
 \right),
\end{eqnarray}
where $\bar{\boldsymbol \Lambda}$ is an $L\times L$ diagonal matrix with diagonal elements $[\alpha_1^2,\alpha_2^2,\ldots,\alpha_{L}^2]$, $\Re\{\cdot\}$ standing for the real part of a complex number.
The first term of the MSE expression is:
\begin{eqnarray}
\label{eqn_mse_h2_t1}
&&  \bar{\bar{\boldsymbol \Delta}} \bar{\boldsymbol \Lambda} (\bar{\bar{\boldsymbol \Delta}})^{\mathcal{H}}= \nonumber\\
&&\!\!\!\!\!\!\left[ \begin{smallmatrix}
\alpha_1^2\frac{(L-1)^2}{N^4}&+\alpha_2^2\frac{(L-2)^2}{N^4}+ &\cdots+&  \alpha_{L}^2\!\frac{(L-2)^2}{N^4}    &    \\
  &\alpha_1^2\frac{(L-2)^2}{N^4}+ & \alpha_2^2\frac{(L-1)^2}{N^4}+ &\cdots+\alpha_{L}^2\!\frac{(L-2)^2}{N^4}    & \\
  &  & \ddots  &    &    \\
  &       & \alpha_1^2\frac{(L-2)^2}{N^4}+&\alpha_2^2\frac{(L-2)^2}{N^4}+\cdots &+\alpha_{L}^2\frac{(L-1)^2}{N^4} \\
 \end{smallmatrix}\right]\nonumber \\
\end{eqnarray}

The third term of the MSE expression is:
\begin{eqnarray}
  \label{eqn_mse_h2_t3}
 \frac{\sigma_w^2}{N} \bar{\boldsymbol \Delta} \bar{\textbf{Q}}\bar{\boldsymbol \Delta}^{\mathcal{H}}= \frac{(L-1)(N-L+2)\sigma_w^2}{N^4} \textbf{I}_{L}.
\end{eqnarray}

The fourth term of the MSE expression is:
\begin{eqnarray}
   \label{eqn_mse_h2_t4}
   &&\frac{\sigma_w^2}{N}\bar{\textbf{Q}}\bar{\boldsymbol \Delta}^{\mathcal{H}}=\frac{(L-1)\sigma_w^2}{N^3} \textbf{I}_{L}.
\end{eqnarray}
Eventually, putting (\ref{eqn_mse_h2_t1}), (\ref{eqn_mse_h2_t3}) and (\ref{eqn_mse_h2_t4}) into (\ref{eqn_mse_h2_general}) yields:
\begin{equation}
\label{eqn_mse_h2_final}
  \varepsilon_{\hat{h}_3}= \frac{N^3+(L-1)(2-L-N)}{N^4}\sigma_w^2+ \frac{(L-1)(L^2-3L+3)}{N^4L}.%\frac{1}{N}\sigma_w^2+\frac{(L-1)(2-L-N)}{N^4}\sigma_w^2.
\end{equation}
Comparing (\ref{eqn_mse_h2_final}) with (\ref{eqn_mse_bar_h_time}), in low SNR region, the MSE approaches the Cram\'er-Rao bound, while, in high SNR region, the estimation error floor is reduced by a ratio of approximately $\left( \frac{L}{N}\right)^2$.

\section{Complexity Analysis}

\begin{figure}[!t]
\centering
\includegraphics[width=3.4in]{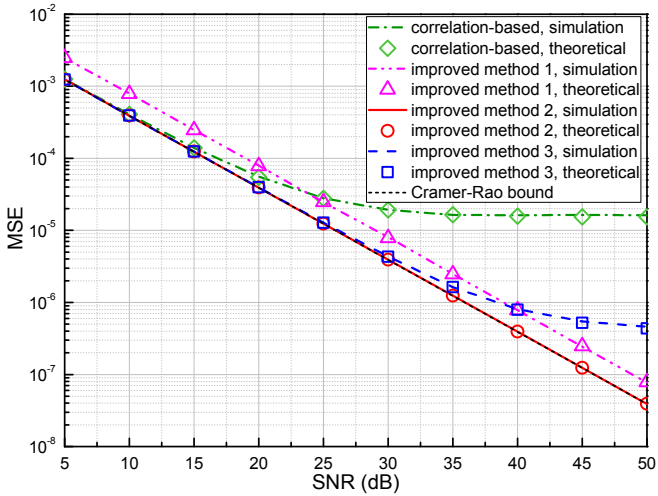}
\caption{MSE performance of different estimators with $\nu=420$, ${N}=255$, ${N_{_{\textrm{CP}}}}=165$ in the TU-6 channel.}
\label{fig_MSE_PN_time_est_420}
\end{figure}

\begin{table}[!t]
\renewcommand{\arraystretch}{1.3}
\caption{Complexities of the PN-based channel estimation methods.}
\label{tbl_complexity_PN_based_methods}
\centering
\begin{tabular}{|m{5cm}||m{1.5cm}|}
\hline
{\textbf{Estimator} }     & \textbf{Complexity}\\ \hline\hline
Circular correlation-based estimator~(\ref{eqn_bar_h_mat}) & $\mathcal{O}$($N^2$) or $\mathcal{O}$($N\cdot\log N$) \\ \hline\hline
Multiplying  matrix inverse $N$-length (\ref{eqn_est_mult_q_inv}) & $\mathcal{O}$($N^2$) \\
Multiplying  matrix inverse $L$-length (\ref{eqn_est_q_inv_trunc})  & $\mathcal{O}$($L^2$)\\
Subtracting interference (\ref{eqn_h_est_subract}) & $\mathcal{O}$($L^2$)\\ \hline
\end{tabular}
\end{table}

In order to compare the computational complexity of each estimation method, we investigate the required basic operations, i.e. multiplication, additions and FFT's etc. Table~\ref{tbl_complexity_PN_based_methods} shows the complexity of the basic correlation-based estimator (\ref{eqn_bar_h_mat}) and the additional complexities required by the improved methods (\ref{eqn_est_mult_q_inv}), (\ref{eqn_est_q_inv_trunc}) and (\ref{eqn_h_est_subract}).

The $N$-length circular convolution takes $N$ multiplications and $N-1$ additions for each delay. That is $N^2$ multiplications and $N\times (N-1)$ additions for an $N$-length CIR estimate. Therefore, the complexity of the circular correlation-based estimator (\ref{eqn_bar_h_mat}) is $\mathcal{O}$($N^2$). Considering that the circular convolution can also be computed by using the FFT, the computational complexity can be reduced to $\mathcal{O}$($N\cdot\log N$).

As far as the improved estimators are concerned, additional complexities are needed by the estimation refinement process. For instance, in the estimator (\ref{eqn_est_mult_q_inv}), the matrix $\textbf{Q}^{-1}$ depends only on the given PN sequence and can thus be pre-computed and stored~\footnote{In fact, the matrix $\textbf{Q}^{-1}$ is quite structured. There are only two values -- diagonal and off-diagonal elements. Hence, it is not necessary to store all the elements of the matrix. Instead, it is smarter to record the two values only. The cost of the storage is therefore negligible.}. The matrix multiplication needs $N^2$ multiplications and $N\times(N-1)$ additions. Therefore, the additional complexity is $\mathcal{O}$($N^2$).
In the meantime, the matrix multiplication can be carried out in a reduced length $L$ as done in estimator (\ref{eqn_est_q_inv_trunc}). For the same reason as in the pervious situation, the computation of $\bar{\textbf{Q}}^{-1}$ does not need any additional effort. The corresponding additional complexity is therefore reduced to $\mathcal{O}$($L^2$).
For the estimator (\ref{eqn_h_est_subract}), the refinement of each channel tap needs ($L-1$) additions. That is, $L\times(L-1)$ additions are required for the whole CIR estimate, which corresponds to a computational complexity of $\mathcal{O}$($L^2$).

\begin{figure}[!t]
\centering
\includegraphics[width=3.4in]{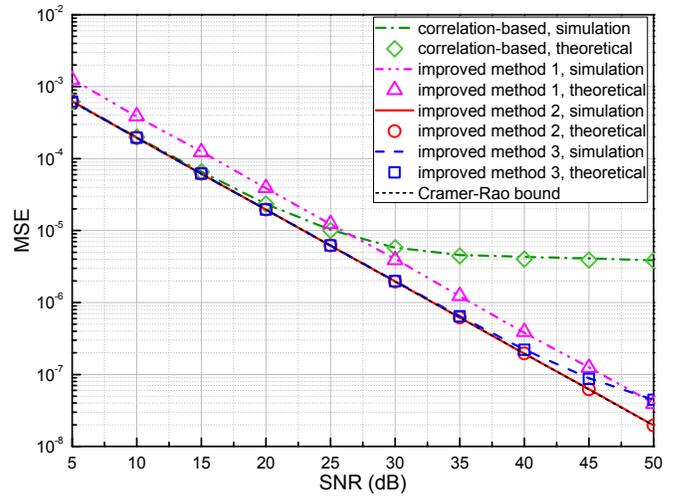}
\caption{MSE performance of different estimators with $\nu=945$, ${N}=511$, ${N_{_{\textrm{CP}}}}=434$ in the TU-6 channel.}
\label{fig_MSE_PN_time_est_945}
\end{figure}

\begin{table}[t]
 \begin{center}
 \caption{Profile of the TU-6 and HT channels}
 \label{TU6}
 {\renewcommand\arraystretch{1.3}
   \begin{tabular}{|c|m{1.5cm}||m{0.5cm}|m{0.5cm}|m{0.5cm}|m{0.5cm}|m{0.5cm}|m{0.5cm}|}
        \hline
        \multicolumn{2}{|c||}{\textbf{Channel}} & \small \textbf{Tap1} & \small \textbf{Tap2} & \small \textbf{Tap3}& \small\textbf{Tap4}& \small \textbf{Tap5} & \small \textbf{Tap6}\\
        \hline \hline
        \multirow{2}{0.3cm}{TU} & \small Delay ($\mu s$) &\small 0 &\small 0.2 &\small 0.5 &\small 1.6 &\small 2.3 &\small 5.0 \\
        \cline{2-8}
         & \small Power (dB) &\small -3 &\small 0 &\small -5 &\small -6 &\small -8 &\small -10 \\
        \hline\hline
        \multirow{2}{0.3cm}{HT} & \small Delay ($\mu s$) &\small 0 &\small 0.2 &\small 0.4 &\small 0.6 &\small 15.0 &\small 17.2 \\
        \cline{2-8}
         & \small Power (dB) &\small 0 &\small -2 &\small -4 &\small -7 &\small -6 &\small -12 \\      \hline
    \end{tabular}}
 \end{center}
\end{table}

\section{Simulation Results}
\begin{figure}[!t]
\centering
\includegraphics[width=3.4in]{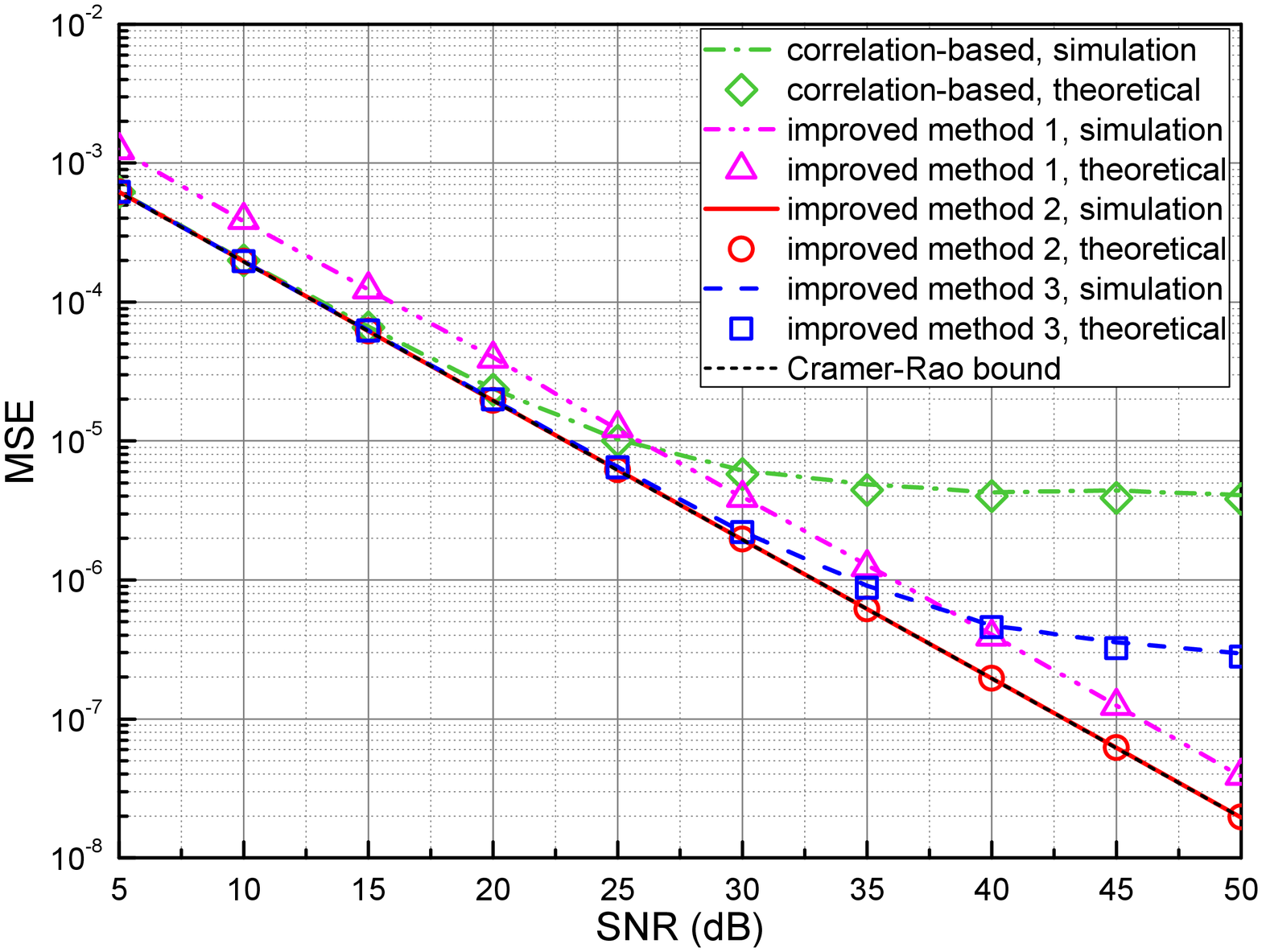}
\caption{MSE performance of different estimators with $\nu=945$, ${N}=511$, ${N_{_{\textrm{CP}}}}=434$ in the HT channel.}
\label{fig_MSE_PN_time_est_945_HT}
\end{figure}

The simulation parameters are selected according to the specifications of the DTMB standard~~\cite{DTMB_Standard} where the sampling period is $1/7.56\ \mu$s. The PN sequences are generated using the maximal linear feedback shift registers specified by the standard as well.
The Typical Urban with six paths (TU-6) and Hilly Terrain (HT) channels specified in~\cite{failli1988cost} are used in the evaluation. The power delay profiles of the two channels are given in Table~\ref{TU6}.
The maximum delays of the TU-6 channel and HT channel are 5 $\mu$s and 17.2 $\mu$s which correspond to $L=38$ and $L=130$ samples of the DTMB system, respectively.
The performance of the different estimators is investigated with different PN lengths and channels as shown from Figure~\ref{fig_MSE_PN_time_est_420} to Figure~\ref{fig_MSE_PN_time_est_945_HT}.
Results are obtained from 1000 realizations of each channel.
The correlation-based estimator corresponds to the classical approach found in the literature~\cite{wang2005iterative} while others are the improved versions introduced in section~\ref{section:improved_time_est}.A, B and C.
From the figures, we can have the following observations:

\begin{figure}[!t]
\centering
\includegraphics[width=3.4in]{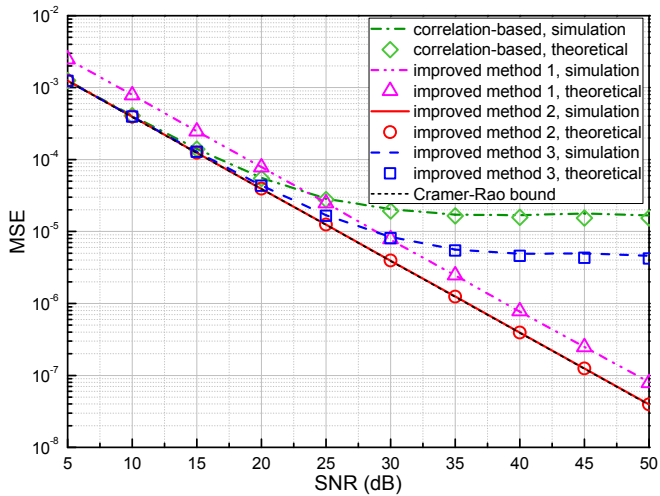}
\caption{MSE performance of different estimators with $\nu=420$, ${N}=255$, ${N_{_{\textrm{CP}}}}=165$ in the HT channel.}
\label{fig_MSE_PN_time_est_420_HT}
\end{figure}

1.  \textit{Methods {\footnotesize `\textsf{improved method 1}'} and {\footnotesize `\textsf{improved method 2}'} completely eliminate the estimation error floor, while method {\footnotesize `\textsf{improved method 3}'} reduces the error floor when the channel delay spread is shorter than the PN length}. Multiplying the basic PN-correlation-based estimator by the inverse of the correlation matrix can create a perfect identity matrix which leads to estimators free of estimation error floor. This can be seen from the performance of {\footnotesize `\textsf{improved method 1}'} and {\footnotesize `\textsf{improved method 2}'}.
On the other hand, {\footnotesize `\textsf{improved method 3}'} subtracts the interference components using the CIR estimates.
It approaches the Cram\'er-Rao bound in low SNR region, while the estimation error floor is reduced by a ratio of approximately $(\frac{L}{N})^2$ in high SNR region. It indicates that the reduction is more notable when the channel delay is significantly small compared to the PN length.
For instance, comparing Fig.~\ref{fig_MSE_PN_time_est_420} and Fig.~\ref{fig_MSE_PN_time_est_420_HT}, {\footnotesize `\textsf{improved method 3}'} obtains more improvement in the TU-6 channel than in the HT channel given the same PN.
In addition, {\footnotesize `\textsf{improved method 3}'} can achieve a lower estimation error floor when a longer PN is used (comparing Fig.~\ref{fig_MSE_PN_time_est_420} and Fig.~\ref{fig_MSE_PN_time_est_945}, or Fig.~\ref{fig_MSE_PN_time_est_420_HT} and Fig.~\ref{fig_MSE_PN_time_est_945_HT}).

2. \textit{Method {\footnotesize `\textsf{improved method 1}'} boosts the noise variance, while methods {\footnotesize `\textsf{improved method 2}'} and {\footnotesize `\textsf{improved method 3}'} do not.}
From this point, the latter two methods are the preferred ones because they outperform the classical correlation based approach whatever the SNR. In contrast, {\footnotesize `\textsf{improved method 1}'} presents a performance back-off due to the noise component power boost, which leads to worse MSE results than the correlation based approach at low SNR.
Last but not least, the estimator {\footnotesize `\textsf{improved method 2}'} does not have estimation error floor and achieves the Cram\'er-Rao bound.

\section{Conclusion}
In this paper, we have investigated the PN-correlation-based channel estimator for TDS-OFDM.
Aiming at reducing the estimation error floor encountered by the classical  PN-correlation-based estimator, we have proposed three improved estimators which exploit the correlation property of the m-sequence and the knowledge of the channel length.
It has been shown through mathematical derivations and simulations that the new proposed estimators approaches or even achieve the Cram\'er-Rao bound.

\section*{Acknowledgment}
The authors would like to thank the European CELTIC project ``ENGINES'' for its support of this work.

%\balance

\ifCLASSOPTIONcaptionsoff
  \newpage
\fi

\bibliographystyle{IEEEtran}
\bibliography{letters}
\end{document}